\pdfoutput=1
\documentclass[aps,prd,twocolumn,superscriptaddress,showpacs,tightenlines,nofootinbib]{revtex4}
\usepackage{graphicx}
\usepackage{epsfig}
\usepackage{bm}
\usepackage{latexsym,amssymb,amsmath,amsfonts,amssymb,txfonts,pxfonts,wasysym,float}
\usepackage{color}

\newcommand{\postscript}[2]{\setlength{\epsfxsize}{#2\hsize}
   \centerline{\epsfbox{#1}}}

\usepackage[usenames,dvipsnames]{xcolor}
\definecolor{orange}{cmyk}{0,0.5,1,0}
\definecolor{rossoCP3}{cmyk}{0,.88,.77,.40}
\definecolor{graa}{rgb}{0.8,0.8,0.8}
\definecolor{blaa}{rgb}{0.2,0.2,0.6}

\begin{document}

\title{\color{rossoCP3}{Strange fireball as an explanation of the muon excess in Auger data}}

\author{Luis A. Anchordoqui}
\affiliation{Department of Physics \& Astronomy,  Lehman College, City University of
  New York, NY 10468, USA}
\affiliation{Department of Physics,
 Graduate Center, City University
  of New York,  NY 10016, USA}
\affiliation{Department of Astrophysics,
 American Museum of Natural History, NY
 10024, USA}

\author{Haim Goldberg}
\affiliation{Department of Physics,
Northeastern University, Boston, MA 02115, USA}

\author{Thomas J. Weiler}
\affiliation{Department of Physics \& Astronomy, Vanderbilt University, Nashville, TN 37235, USA}

\begin{abstract}
  \vskip 2mm \noindent We argue that ultrahigh energy cosmic ray
  collisions in the Earth's atmosphere can probe the strange quark
  density of the nucleon. These collisions have center-of-mass
  energies $\agt 10^{4.6} \, A \, {\rm GeV}$, where $A \geq 14$ is the
  nuclear baryon number. We hypothesize  the formation of a deconfined
  thermal fireball which undergoes a sudden hadronization.  At
  production the fireball has a very high matter density and consists
  of gluons and two flavors of light quarks ($u,d$). Because the
  fireball is formed in the baryon-rich projectile fragmentation
  region, the high baryochemical potential damps the production of $u
  \bar u$ and $d \bar d$ pairs, resulting in gluon fragmentation
  mainly into $s \bar s$. The strange quarks then become much more
  abundant and upon hadronization the relative density of strange
  hadrons is significantly enhanced over that resulting from a hadron
  gas.  Assuming the momentum distribution functions can be
  approximated by Fermi-Dirac and Bose-Einstein statistics, we
  estimate a kaon-to-pion ratio of about $3$ and expect a similar
 (total) baryon-to-pion ratio.  We show that, if this were the case,
  the excess of strange hadrons would suppress the fraction of energy
  which is transferred to decaying $\pi^0$'s by about 20\%, yielding a
  $\sim 40\%$ enhancement of the muon content in atmospheric cascades,
  in agreement with recent data reported by the Pierre Auger
  Collaboration.
\end{abstract}
\date{December 2016}
\maketitle

\section{Introduction}

Ultrahigh-energy ($E \agt 10^{9.8}~{\rm GeV}$) cosmic rays provide a
formidable beam to study particle collisions at center-of-mass
energies and kinematic regimes not accessible at terrestrial
accelerators. The incident cosmic radiation interacts with the atomic
nuclei of air molecules and produces air showers which spread out over
large areas. If the primary cosmic ray is a baryon, hundreds to
thousands of secondary particles are usually produced at the
interaction vertex, many of which also have energies above the highest
accelerator energies~\cite{Anchordoqui:1998nq}. These secondary
products are of course intrinsically hadrons. Generally speaking, by
extrapolating final states observed at collider experiments, we can
infer that, for $pp$ collisions at center-of-mass energy $\sqrt{s}
\sim 140~{\rm TeV}$, the jet of hadrons contains about 75\% pions
(including 25\% $\pi^0$'s, in accord with isospin invariance), 15\%
kaons, and 10\% nucleons~\cite{GarciaCanal:2009xq}.

During the shower evolution, the hadrons propagate through a medium
with an increasing density as the altitude decreases and the
hadron-air cross section rises slowly with energy. Therefore, the
probability for interacting with air before decay increases with
rising energy. Moreover, the relativistic time dilation increases the
decay length by a factor $E_h/m_h$, where $E_h$ and $m_h$ are the
energy and mass of the produced hadron.  When the $\pi^0$'s (with a
lifetime of $\simeq 8.4 \times10^{-17}~{\rm s}$) do decay promptly to
two photons, they feed the electromagnetic component of the
shower. For other longer-lived mesons, it is instructive to estimate
the critical energy at which the chances for interaction and decay are
equal. For a vertical transversal of the atmosphere, such a critical
energy is found to be: $\xi_c^{\pi^\pm} \sim 115~{\rm GeV}$,
$\xi_c^{K^\pm} \sim 850~{\rm GeV}$, $\xi_c^{K^0_L} \sim 210~{\rm
  GeV}$, $\xi_c^{K^0_S} \sim 30~{\rm TeV}$~\cite{Gondolo:1995fq}.  The
dominant $K^+$ branching ratios are to $\mu^+ \nu_\mu\ (64\%)$, to
$\pi^+ \pi^0\ (21\%)$, to $\pi^+\pi^+\pi^-\ (6\%)$, and to $\pi^+
\pi^0\pi^0\ (2\%)$, whereas those of the $K^0_S$ are to $\pi^+\pi^- \
(60\%)$, to $\pi^0 \pi^0 \ (30\%)$, and for $K^0_L$ we have $\pi^\pm
e^\mp \nu_e \ (40\%)$, $\pi^\pm \mu^\mp \nu_\mu \ (27\%)$, $\pi^0
\pi^0 \pi^0 \ (19\%)$, $\pi^+ \pi^- \pi^0 \
(12\%)$~\cite{Olive:2016xmw}.  With these figures in mind, to a first
approximation it seems reasonable to assume that in each generation of
particles about 25\% of the energy is transferred to the
electromagnetic shower, and all hadrons with
energy~$\agt\xi_c^{\pi^\pm}$ interact rather than decay, continuing
to produce the hadronic shower.\footnote{The electromagnetic
shower fraction from pions only is less than 25\%, but simulations show that inclusion of
  other hadronic resonances brings the electromagnetic shower fraction
  up to about 25\%~\cite{Aab:2016hkv}. We take 25\% as a reasonable
  estimate of the energy transfer to the electromagnetic shower.}
Eventually, the electromagnetic cascade dissipates around 90\% of the
primary particle's energy and the remaining 10\% is carried by muons
and neutrinos.

As the cascade process develops in the atmosphere, the number of
particles in the shower increases until the energy of the secondary
particles is degraded to the level where ionization losses
dominate. At this point the density of particles starts to
decline. The number of particles as a function of the amount of
atmosphere penetrated by the cascade ($X$ in ${\rm g \, cm}^{-2}$) is
known as the longitudinal profile.  A well-defined peak in the
longitudinal development, $X_{\rm max}$, occurs where the number of $e^\pm$ in the
electromagnetic shower is at its maximum. $X_{\rm max}$ increases with
primary energy, as more cascade generations are required to degrade
the secondary particle energies. Evaluating $X_{\rm max}$ is a
fundamental part of many of the composition analyses done when
studying air showers. The generic shower properties can be
qualitatively well understood using the superposition principle, which
states that a shower initiated by a nucleus with $A$ nucleons and
energy $E$ behaves to a good approximation as the superposition of $A$
proton showers with initial energy
$E/A$~\cite{Anchordoqui:2004xb}. This phenomenological assumption
relies on the fact that the effect of nuclear
binding must be negligible at extremely high energies. Thus, for a given total energy $E$, showers
initiated by a heavy nucleus have smaller $X_{\rm max}$ than proton
induced showers.

The integrated
longitudinal profile provides a calorimetric measurement of the energy
of the primary cosmic ray, with a relatively small uncertainty due to
the correction for energy lost to neutrinos and particles hitting the
ground. The characteristics of the cascade depend dominantly on the
elasticity (fraction of incoming energy carried by the leading
secondary particle) and  the multiplicity of secondary particles in the
early, high-energy interactions. Modeling the development of a cosmic-ray air shower  requires
extrapolation of hadronic interaction models tuned to accommodate LHC
data~\cite{d'Enterria:2011kw}. Not surprisingly, such extrapolation usually leads to
discrepancies between measured and simulated shower properties. The
hadronic interaction models are further constrained by independent
measurements of  $X_{\rm max}$ and the density of muons at 
$1~{\rm  km}$ from the shower core $N_\mu$~\cite{Ulrich:2010rg}.  The mean $X_{\rm max}$
is primarily sensitive to the cross-section, elasticity, multiplicity,
and primary mass.  The mean $N_\mu$ is primarily sensitive to the
multiplicity, the $\pi^0$ energy fraction (the fraction of incident
energy carried by $\pi^0$'s in hadronic interactions), and the primary
mass. 

Over the past few decades, it has been suspected that the number of
registered muons at the surface of the Earth is by some tens of
percentage points higher than expected with extrapolations of existing
hadronic interaction
models~\cite{AbuZayyad:1999xa,Aab:2014pza}.\footnote{See, however,
  \cite{Fomin:2016kul}.}  Very recently, a study from the Pierre Auger
Collaboration~\cite{ThePierreAuger:2015rma} has strengthened this
suspicion, using a novel technique to mitigate some of the measurement
uncertainties of earlier methods~\cite{Aab:2016hkv}. The new analysis
of Auger data suggests that the hadronic component of showers (with
primary energy $10^{9.8} < E/{\rm GeV} < 10^{10.2}$) contains about
$30\%$ to $60\%$ more muons than expected. The significance of the
discrepancy between data and model prediction is somewhat above $2.1
\sigma$.

Changing the $\pi^0$ energy fraction or suppressing $\pi^0$ decay are
the only modifications which can be used to increase $N_\mu$ without
coming into conflict with the $X_{\rm max}$
observations~\cite{Farrar:2013sfa}.\footnote{We note in passing that
  muons may be able to escape the shower core before reaching the
  Earth surface if their production angle is increased by boosting the
  $p_T$ distribution. Since the muon lateral distribution function is
  a steeply falling function of the radius, a larger production angle
  would increase $N_\mu$. However, as shown in~\cite{Farrar:2013sfa},
  the measured zenith angle dependence of the ground signal vetoes the
  correlated flattening of the muon lateral distribution function
  necessary to accommodate Auger data.} Several new physics models
have been proposed exploring these two
possibilities~\cite{Farrar:2013sfa,Allen:2013hfa,AlvarezMuniz:2012dd}. In
this paper we adopt a purely phenomenological approach to develop an
alternative scheme. In sharp contrast to previous models, our proposal
is based on the assumption that ultrahigh energy cosmic rays are heavy
(or medium mass) nuclei. Our work builds upon some established
concepts, yet contravenes others.

We conceive the production and separation of strangeness in a
baryon-rich {\it Centauro-like} fireball before its spontaneous {\it
  explosive}
decay~\cite{Halzen:1985eh,Panagiotou:1989zz,Panagiotou:1991kv,Asprouli:1994ch,Angelis:2003zn}. At
production the fireball has a very high matter density and consists of
gluons and two flavors of light quarks ($u,d$). Because the fireball
is formed in the baryon-rich projectile fragmentation region, the high
baryochemical potential slows down the creation of $u \bar u$ and $d
\bar d$ pairs, resulting in gluon fragmentation dominated by $ g \to s
\bar s$. The larger amount of $u$ and $d$ with respect to $\bar u$ and
$\bar d$ gives a higher probability for $\bar s$ to find $u$ or $d$
and form $K^+$ or $K^0$ than for $s$ to form the antiparticle
counterparts.  Prompt {\it hard} kaon emission then carries away all
strange antiquarks and positive charge, lowering somewhat the initial
temperature and entropy.  The late-stage hadronization is
characterized by production of $K^-$, ${\bar K}^0$, nucleons, pions,
and strange baryons. Overall, after hadronization is complete, the
relative density of strange hadrons is significantly enhanced over
that resulting from a hadron gas alone, damping the $\pi^0$ energy
fraction.

The layout of the paper is as follows. We begin in Sec.~\ref{s2} with
an overview of the fireball paradigm and make a critical review of the
available experimental data from colliders. After that, in
Sec.~\ref{s3} we discuss the particulars of air shower evolution and
present our results from Monte Carlo simulations. We show that the
formation of a plasma, with gluons and massive quarks, could play a
key role in the hadronization process, modifying shower observables.
In particular, we demonstrate that air showers triggered by a fireball
explosion tend to increase $N_\mu$, and under some reasonable
assumptions can accommodate Auger data. Finally, we summarize our
results and draw our conclusions in Sec.~\ref{s3}.

\section{Fireball Phenomenology}
\label{s2}

It has long been suspected that, for systems of high energy density,
the elementary excitations can be safely approximated by an ensemble
of free quarks and gluons at finite temperature and baryon number
density~\cite{Collins:1974ky,Iachello:1974vx,Cabibbo:1975ig}. 
This is because when the energy density is extremely high,
the expected average particle separation is so small that the
effective strength of interactions is
weak (asymptotic freedom)~\cite{Gross:1973id,Politzer:1973fx}.
For many purposes, the
order of the energy density of matter inside a heavy nucleus is
immense, 
\begin{equation} 
\varepsilon_A \sim \frac{m_p A}{\frac{4}{3} \pi R^3} \sim 0.15~{\rm GeV/fm^3} 
\end{equation} 
where $m_p$ is the proton mass, $A$ is the nuclear baryon number, and
$R \sim 1.1 A^{1/3}~{\rm fm}$ is the nuclear radius. However, at
typical energy densities inside of nuclei, quarks and gluons are very
probably confined, on the average, inside of hadrons such as protons,
neutrons, or pions.  For $\varepsilon \gg 0.15~{\rm GeV/fm^3}$ all of
these hadrons could be squeezed so tightly together that on the
average they will all overlap and the system would become an
unconfined plasma of quarks and gluons, which are free to roam the
system. The energy scale at which hadrons begin to overlap is above
the energy density of matter inside the proton, $\varepsilon_p \sim
0.5~{\rm GeV/fm^3}$, where we have taken the radius of the proton as
measured in electron scattering $R \sim 0.8~{\rm fm}$. Indeed, using
phenomenological considerations it is straightforward to estimate that
the critical energy density to form a non-hadronic medium is around
$1~{\rm GeV/fm^3}$~\cite{Bjorken:1982qr}. This result is supported by
high statistics lattice-QCD calculations, which yield $\varepsilon_c
\sim 7T_c^4 \sim 1~{\rm GeV/fm}^3$, where we have taken $T_c =
190~{\rm MeV}$~\cite{Bazavov:2009zn}.

Besides the early universe, the conditions of extremely high
temperature and density necessary for the appearance of unconfined
quark and gluons could occur in at least two other physical phenomena:
{\it (i)}~the interiors of neutron
stars~\cite{Ivanenko:1969gs,Itoh:1970uw,Chapline:1976gq,Freedman:1977gz,Chapline:1977rn,Alcock:1986hz,Olinto:1986je}
and {\it (ii)}~high-energy nucleus-nucleus collisions, whether
artificially produced at accelerators or naturally occurring
interactions of cosmic rays with particles in the Earth's
atmosphere~\cite{Anishetty:1980zp,Halzen:1981zx,Cleymans:1982cc,Halzen:1983gx}. We
estimate the energy density in nucleus-nucleus collisions of cosmic
rays following~\cite{Bjorken:1982qr}. We assume there exists a
``central-plateau'' structure in the inclusive particle productions as
function of the pseudorapidity variable. The energy per particle
should be of the order of the typical transverse momentum per particle
$\langle p_T \rangle$. More precisely, in the fireball frame each
isotropically emitted particle has an energy given by
\begin{equation}
\langle E \rangle \sim \left[(4 \langle p_T\rangle /\pi)^2 + m^2
\right]^{1/2} \, ,
\end{equation}
where $m$ is the particle's mass~\cite{Panagiotou:1991kv}.  The energy
content is approximately $\delta_A \langle E \rangle dN_{\rm ch}$,
where $dN_{\rm ch}$ is the charged-hadron multiplicity per $pp$
collision, and $\delta_A$ the number of nucleon-nucleon interactions
in the fireball during the collision.  Consider two nuclei of
transverse radius $R$ which collide in the center-of-mass frame. The
longitudinal size of the nuclei is Lorentz contracted forming a
transverse thin slab at mid-pseudorapidity. The initial {\it fireball}
volume is $dV = (R^2 \pi) \tau_0 d\eta_0$, where $\tau_0$ is the
typical time scale for the formation and decay of a central fireball
and hence $\tau_0 d\eta_0$ is the longitudinal size, with $d\eta_0$
the pseudorapidity width at $\tau_0$. The time scale can be estimated
as the time to traverse at light speed the fireball diameter, i.e.,
$\tau_0 = 2R/c \sim 1~{\rm fm}/c$~\cite{Apolinario:2017sob}.  The energy density during
the cosmic ray (CR) collision is then
\begin{equation}
\varepsilon_{\rm CR} \sim \delta_A \ \langle
E \rangle  \ \frac{1}{\pi R^2 \tau_0} \ \left. \frac{dN_h}{d\eta}
\right|_{\eta = 0} \, . 
\end{equation}
The average transverse-momentum of charged hadrons produced in $pp$
collisions can be parametrized as a function of the
squared center-of-mass energy,
\begin{equation}
\langle p_T \rangle = \left(0.413 - 0.0171 \ln s  + 0.00143
\ln^2 s \right)~{\rm GeV}  \,,
\end{equation}          
with $s$ in appropriate units of ${\rm GeV}^2$~\cite{Khachatryan:2010us}.  
On the other hand, for a cosmic-ray nitrogen nucleus 
(for simplicity, we choose the beam nucleus to be that
of the dominant element in air\footnote{In air, nitrogen is a dimer, N$_2$, with total $A = 2\times 14=28.$
However, the electronic binding is $\sim$~eV, whereas the nuclear binding is $\sim $~MeV, 
so the energy scales and length scales differ by $\sim$~a million, 
and the dimerization is not expected to survive the cosmic-ray production process.
The nuclear binding of the single nucleus apparently does survive this process.
})
of $E \approx 10^{10}~{\rm GeV}$ colliding with an air
nucleus, we have 
$\sqrt{s} \approx 10^{4.6} A~{\rm GeV}$, yielding $ \langle p_T
\rangle \sim 0.69~{\rm GeV}$ and $(dN_{\rm ch}/d\eta)_{\eta =0} \sim
7$~\cite{d'Enterria:2011jc}. Throughout, we take $A =14$ as fiducial in
our calculations and assume that half of these number of nucleons interact
per collision, producing the fireball. Therefore, taking an effective mass $m \sim 500~{\rm
  MeV}$ and $\delta_A = A/2$, the energy density in such a scattering
process is
\begin{equation}
\varepsilon_{\rm CR}
\sim 2.2~{\rm GeV/fm^3} \,,
\label{ecr}
\end{equation}
well above $\varepsilon_c$, complying with
the requirement for the formation of a deconfined thermal fireball. 

We envision the fireball as a plasma of {\it massive} quarks and gluons maintained in
both kinetic and chemical equilibrium.  Because the total number of
particles is allowed to fluctuate, we adopt the viewpoint of the grand
canonical ensemble.  In this representation, which allows exchange of
particles among the system and the reservoir, the control
variables are the baryochemical potential $\mu_B$ and the temperature
$T$. In the limit $\mu_B \to 0$ and $T \to 0$ the system becomes the vacuum.

The momentum distribution functions $f(p)$ can be approximated by
Fermi-Dirac $(+)$ and Bose-Einstein $(-)$ statistics,
\begin{equation}
f_{i,\pm} (p) = \left[\exp \left( \frac{\sqrt{p^2 + m_i^2}-\mu_i}{T}
  \right) \pm 1 \right]^{-1}
\,,
\label{FD-BE}
\end{equation}
yielding following equilibrium number densities  
\begin{equation}
n_{i,\pm}  =  g_i \int \frac{d^3p}{(2 \pi)^3} \, f_{i,\pm}  (p) \,,
\label{numberD}
\end{equation}
where the index $i$ runs over $\{u,d,s,g\}$, $\mu_s = 0$ because of the total strangeness
neutrality of the initial state,  \mbox{$\mu_u = \mu_d \equiv
\mu_q = \mu_B/3$,} $p$ and $m_i$ are the particle's
momentum and mass, and $g_i$ is the spin-color degeneracy factor. The
plus sign is to be used for quarks and minus sign for gluons, with
$\mu_g =0$. The density of the strange quarks is found to be (two
spins and three colors)
\begin{eqnarray}
n_s = n_{\bar s} & = & 6 \int \frac{d^3p}{(2 \pi)^3} \
\frac{1}{e^{\sqrt{p^2 + m_s^2}/T} + 1} \nonumber \\
& = & \frac{3}{\pi^2} T^3
  \sum_{n=1}^\infty \frac{(-1)^{n+1}}{n^3} (n m_s/T)^2 \ K_2 (n m_s/T) \nonumber \\
& \approx & \frac{3}{\pi^2} m_s^2 T \  K_2 (m_s/T) \,, 
\label{ns}
\end{eqnarray}
where $K_2(x)$ is the second order modified Bessel
function~\cite{Rafelski:1983hg,Glendenning:1984ta}.  
In addition, there is a certain light antiquark density ($\bar q$ stands
for either $\bar u$ or $\bar d$):
\begin{eqnarray}
n_{\bar q} & = &  6 \int \frac{d^3p}{(2 \pi)^3} \
\frac{1}{e^{|p|/T + \mu_q/T} + 1} \nonumber \\
& = & \frac{6}{\pi^2}  T^3 \sum _{n=1}^{\infty}
\frac{(-1)^{n-1}}{n^3} e ^{-n \mu_q/T}     \nonumber \\
& \approx & \frac{6}{\pi^2} T^3 e^{-\mu_q/T} \, .
\label{nqbar}
\end{eqnarray}
Note that the baryonic chemical potential  exponentially suppresses
the $q \bar q$ pair production. This reflects the chemical equilibrium
between $q \bar q$ and the presence of a light quark density
associated with the net baryon number. 
Now, since the chemical potentials satisfy $\mu_q = - \mu_{\bar q}$,
it follows that~\cite{Kolb:1990vq} 
\begin{eqnarray}
n_q - n_{\bar q} & = & \frac{g_i}{2 \pi^2} \int_0^\infty dp \, p^2
\left(\frac{1}{e^{(p-\mu_q)/T} +1}
 - \frac{1}{e^{(p + \mu_q)/T} +1} \right)\nonumber \\
& = & \frac{\mu_q^3}{\pi^2} + \mu_q
  T^2  \, .
\end{eqnarray}
Note that this result is exact and not a truncated series.
The gluon density also follows from (\ref{numberD}) and is given by
\begin{equation}
n_g = \frac{16}{\pi^2} \zeta(3) T^3 \, ,
\label{ng}
\end{equation}
where $\zeta(x)$ is the Riemann function. (See Appendix~\ref{App_A}
for details.)  By comparing (\ref{ns}) and (\ref{nqbar}),  it is straightforward to see that there are often more $\bar s$ than anti-quarks of each light flavor,
\begin{equation}
\frac{n_{\bar s}}{n_{\bar q}} \approx \frac{1}{2} \left(\frac{m_s}{T}
\right) \ K_2 (m_s/T) \ e^{\mu_q/T} \  .
\label{stoq}
\end{equation} 
For $T \agt m_s$ and $\mu_B \to 0$, there are about as many $u$ and $d$ quarks as there
are $s$ quarks.

For $T \gg T_c$, many of the properties of the quark-gluon plasma can be calculated in the framework of thermal perturbation theory.
Neglecting quark masses in first-order perturbation theory,
the energy density of the {\it ideal} quark gluon plasma  is found to be
\begin{eqnarray}
\varepsilon_{\rm QGP} & = & \left[\frac{1}{30} \left(N_g + \frac{7}{4} N_q
  \right) - \frac{11 \alpha_s}{3\pi} \right] \pi^2 T^4 +
\left(\frac{N_q}{4} - \frac{6 \alpha_s}{\pi} \right) \mu_q^2 T^2
\nonumber \\
& + & \left(\frac{N_q}{8} - \frac{3 \alpha_s}{\pi} \right)
\frac{\mu_q^4}{\pi^2} + B \,,
\label{eideal}
\end{eqnarray}
where $N_g = 16$ and $N_f$ are the gluon and quark
degrees of freedom, $\alpha_s$ is the QCD coupling constant, and  $B$ is the difference between the energy density of the
perturbative and the nonperturbative QCD vacuum (the bag
constant)~\cite{Muller:1994rb,Asprouli:1995zk,GladyszDziadus:2001cq}.
One observes that (\ref{eideal}) is essentially the equation of state
of a gas of massless particles with corrections due to QCD
perturbative interactions, which are always negative and thus
reduce the energy density at given temperature $T$. 

Since the $u$ and $d$ flavors are almost massless even in the fireball
phase, we can get an estimate of the baryochemical potential by
considering only the contribution from the nearly massless (even in
the fireball phase) two quarks at $T =0$.  For two quark flavors, $N_f
= 12$ and so (\ref{eideal}) simplifies to
\begin{equation}
\varepsilon_{\rm fb}^{T=0} = \left(\frac{3}{2} - \frac{3
    \alpha_s}{\pi}\right)  \  \frac{\mu_q^4}{\pi^2} + B \, .
\label{eidealT0}
\end{equation}
Substituting
(\ref{ecr}) into (\ref{eidealT0}), with $\alpha_s \simeq
0.2$~\cite{Kaczmarek:2005ui} and the MIT bag constant $B^{1/4} \approx 328~{\rm MeV}$~\cite{Asprouli:1995zk},   we
obtain 
\begin{equation}
 \mu_B \sim 2.0~{\rm GeV} \, .
\label{muB}
\end{equation}
The temperature of the plasma can be approximated by~\cite{Kolb:1990vq}
\begin{equation}
T = \frac{2}{3} \ \left( \langle E \rangle - m \right) \sim  580~{\rm MeV} \, .
\label{T}
\end{equation}
A point worth noting at this juncture is that the shapes of the $p_T$
spectra are expected to be determined by an interplay between two
effects: the thermal motion of the particles in the fireball and a
pressure-driven radial flow, induced by the fireball expansion.  To
disentangle the two contributions, namely thermal motion and
transverse flow, one has to rely on model calculations which seem to
indicate that the observed temperature in the particle spectra is
close if not exactly equal to the temperature value that would be
present in the chemically equilibrated fireball~\cite{Rafelski:1996hf}.

Substituting (\ref{muB}) and (\ref{T}) into (\ref{stoq}), with $m_s \simeq
175~{\rm GeV}$, we obtain
$n_{\bar s}/n_{\bar q} \sim 3.1$.  The pion-to-nucleon density ratio
is found to be
\begin{equation}
\frac{n_\pi}{n_N} \sim \varkappa \frac{3}{2} \exp \left\{[m_N -4
  \mu_B/3 - m_\pi]/T  \right\} \,,
\label{pin}
\end{equation}
where the factor 3/2 comes from the number of the particle species,
$m_\pi$ is the pion mass, $m_N \sim 1~{\rm GeV}$ is the average baryon
mass, and $\varkappa$ is a normalization constant fixed by the choice
of the boundary values~\cite{Panagiotou:1991kv}. 

It is worth commenting on an aspect of this analysis which may seem
discrepant at first blush. From relativistic heavy ion experiments one
infers a temperature falling with the chemical potential, from
$T(\mu_B = 0) = 166~{\rm MeV}$. 
However, to accommodate the muon excess
in Auger data one needs $T(\mu_B = 2~{\rm GeV}) = 580~{\rm
  MeV}$. Details of this discrepancy and eventual accommodation are
given in Appendix~\ref{App_ref}.

In our analysis we will adopt a pragmatic approach and avoid the
details of theoretical modeling of the hadronization process.  Which
of the two points of view one may find more convincing, it seems most
conservative at this point to depend on experiment to resolve the
issue. The multiplicity ratio $\pi:K:N \sim 0.15:0.45:0.40$ has been eyeball-fitted to
reproduce the anomalous muon signal observed in Auger data.
We show the success of this protocol in the next section. 
Here we simply note that these ratios are in partial agreement with (\ref{stoq}) and (\ref{pin}) 
for $\varkappa \sim 7$.

\section{Air Shower Evolution}
\label{s3}

We now make contact with the shower evolution. As a first-order
approximation we adopt a basic phenomenological approach. Namely, we
assume that the hadronic shower carries a
fraction $f_h$ of the total energy of the primary cosmic ray $E$,
which scales as
\begin{equation}
f_h  \sim \left( 1 - f_{\rm EM}  \right)^{n_{\rm gen}}   \,,
\label{fh1}
\end{equation}
where $n_{\rm gen}$ is the number of generations required for most pions
to have energies below $\xi_c^{\pi^\pm}$ and $f_{\rm EM}$ is the
average fraction in electromagnetic particles per
generation~\cite{Matthews:2005sd}.  In the canonical framework
hadronic interaction models transfer about 25\% of the energy to the
electromagnetic shower. Conspicuously, the production of light hadrons
in this canonical framework is virtually local in rapidity and,
therefore, since the interaction models are tuned to fit collider data,
$f_{\rm EM}$ would remain approximately constant with energy.  We have
found that for the considerations in the present work we can safely
approximate that each interaction diverts about 75\% of the available
energy into pions, 15\% into kaons, and 10\% continues as
nucleons. Roughly speaking, this is consistent with a multiplicity ratio $\pi:K:N \simeq
0.75:0.15:0.10$. Now, taking then $f_{\rm EM} \sim 0.25$ and $f_h \sim
0.10$ we conclude that $n_{\rm gen} \sim 8$.\footnote{The average
  neutrino energy from the direct pion decay is $\langle E_\nu \rangle
  = (1-r) E_\pi/2 \simeq 0.22 E_\pi$ and that of the muon is $\langle
  E_\mu \rangle = (1+r) E_\pi/2 \simeq 0.78 E_\pi$, where 
  $r=0.573$  is the
  ratio of muon to the pion mass squared. Thus, we can safely neglect
  the missing energy in neutrinos.} This is in agreement with the
estimates of~\cite{Matthews:2005sd} which indicate that the 
number of interactions needed to reach  $\xi_c^{\pi^\pm}$ is $n_{\rm gen} =
3,4,5,6$ for $E = 10^5, 10^6, 10^7, 10^8~{\rm GeV}$, respectively.

A comprehensive study of the
uncertainties associated with the modeling of hadronic interactions
indicates that a simple reduction of $n_{\rm gen}$ to increase $f_h$ correlates
with $X_{\rm max}$, which becomes too shallow before $N_\mu$
is sufficiently increased to accommodate Auger
data~\cite{Farrar:2013sfa}.

Now, if the shower is initiated in a fireball explosion in the first generation, 
then we have seen that we can approximate the multiplicity ratios by 
$\sim 0.15:0.45:0.40$.  Moreover,  this is a completely inelastic
process, but differs from the usual inelastic processes in that a fireball is
also produced. We assume this fireball creates a higher multiplicity
of particles and to a first approximation equally partitions energy
among the secondaries (thereby negating a large leading particle
effect). The fireball production thus  accelerates the cooling of the
cascade and could reduce the number of generations. We denote with
$n'_{\rm gen}$ the numbers of generations for a shower initiated by a fireball. We may then
assume that $n'_{\rm gen} \sim 7$ or 8 would be enough to reach the critical shower
energy $\xi_c^{\pi^\pm}$. To include the fireball effects we rewrite
(\ref{fh1}) as
\begin{equation}
f_h  \sim \left( 1 - f_{\rm EM}^{\rm fb} \right) \  \left( 1 - f_{\rm EM} \right)^{n'_{\rm gen}-1}   \,,
\label{fh2}
\end{equation}
where $f_{\rm EM}^{\rm fb} \sim 0.05$ is the fraction of
electromagnetic energy emitted by the fireball. We arrived at $f_{\rm
  EM}^{\rm fb}$ from considering the $\pi^0$ fraction, equal to 1/3
the total $\pi$ fraction, as before.  By substituting our fiducial
numbers in (\ref{fh2}) it is straightforward to see that the hadronic
shower is increased by about 30\% if $n'_{\rm gen} =8$ and by about
70\% if $n'_{\rm gen}=7$, in agreement with recent Auger
results~\cite{Aab:2016hkv}.

Of course, (\ref{fh2}) is a dreadful simplification of the shower
evolution. As we have noted before, the shower evolution depends on
primary energy, as well as the elasticity and particle multiplicity
which also depend on $E$.  
Note that all these restrictions have not
been specified explicitly as separate parameters in (\ref{fh2}), but
rather as a combined constant $n'_{\rm gen}$.  Moreover, heavy meson
production must also be taken into consideration when modeling the
shower evolution (distinctively, $\eta$ and $\eta'$  
contribute about 4\% to the electromagnetic cascade). Unfortunately,
it is difficult to estimate accurate parameter values of the shower evolution in a simple
analytical fashion. For such a situation, a full-blown simulation may be
the only practical approach.  

As a second order approximation, we estimate the fireball spectrum and
propagate the particles in the atmosphere using the algorithms of {\sc
  aires} (version 2.1.1)~\cite{Sciutto:1999jh}. Most of the large
multiplicity of observable quanta emitted in the fireball is expected
to come through hadronic jets produced by the quarks. 
We assume that $A/2$ nucleons (each with initial energy $E/A$) produce the
  fireball and that the remaining nucleons scatter inelastically at
  the collision vertex. We further assume that for the nucleons
  producing the fireball, each of the valence quarks interact to give
  dijet final states without any leading particle. We calculate 
the total energy of each jet $E_{\rm jet}$ in the rest frame of the
fireball  from the momentum fraction
  carried by the up and down quarks, using CTEQ6L parton distribution
  functions~\cite{Pumplin:2002vw}.

The precise nature of the fragmentation process is unknown. We adopt
  the quark $\rightarrow$ hadron fragmentation spectrum originally
  suggested by Hill
\begin{eqnarray}
\frac{dN_h}{dx} & \approx & 0.08\,\,\exp\left[2.6\sqrt{\ln(1/x)}\right]
\,\,(1-x)^2 \nonumber \\
 & \times & \left[x \sqrt{\ln(1/x)}\right]^{-1}\,.
\label{c}
\end{eqnarray}
where $x \equiv E_h/E_{\rm jet}$, with $E_h$ the energy of any hadron in the jet.
This is consistent with the so-called ``leading-log QCD'' behavior and
seems to reproduce quite well the multiplicity growth as seen in
colliders experiments~\cite{Hill:1982iq}. 
 $dN_h/dx \, \approx\, (15/16) \, x^{-3/2}\, (1 - x)^2$ provides a
reasonable parametrization of (\ref{c}) for $10^{-3} <x<1$.
And so we set the infrared cutoff to $x_{\rm cut} =10^{-3}$.
The main features of the jet fragmentation process derived from this simplified
parameterization are listed in Table~\ref{table}. 
Using the multiplicity ratios derived in the
previous section and the fractional equivalent energies given in
Table~\ref{table}, we construct the fireball particle spectra. 
More specifically, for each jet, we start first from $x=1$ and integrate down in $x$ until three leading hadron 
particles are obtained.  The resulting interval in $x$ is $\Delta x\equiv (x_2=1) - x_1$, as shown in 
column three, with $x_2$ and $x_1$ values listed in columns two and one, respectively.
We assign one of three species-types to each hadron using the $\pi:K:N$ weights.
The energy fraction of the jet contained in these three hadrons is given in column four, and the 
average energy fraction per each of these three hadrons, denoted $x_{\rm equivalent}$, is given in column five.
Next, we duplicate
the procedure for the remaining hadron species following the splitting of fractional energies,
$\int_{x_1}^{x_2} N_h \, dx$, as given in column 3, and assigning to each hadron in the interval 
the corresponding $x_{\rm equivalent}$  from column 5.  
Note that for each subsequent $\Delta x$ interval, the fractional hadron energy $x_{\rm equivalent}$ is significantly smaller;
this feature allows us to sensibly truncate the process after five intervals at our cutoff value $x_{\rm cut}$.
The mean energy of each of the fifth and final batch of 30 hadrons
produced by wee partons is $\sim 1/100^{\rm th}$ 
of the energy of each hadron produced by large $x$ partons in first batch.
The average particle multiplicity per jet is the sum of column three entries, approximately 54. 
Charge and strangeness conservation are separately imposed by hand.

\begin{table}
\caption{Properties of jet hadronization~\cite{Anchordoqui:2001ei}. Columns one and two define the interval $\Delta x=x_2-x_1$ over which
$N_h$~hadrons are made.  $N_h$ is listed in the third column.  Column four presents the fractional 
energy of the jet contained in these $N_h$ hadrons, and column five gives the mean fractional energy 
per each of these hadrons ($x_{\rm equivalent}$).}
\begin{center}
\label{table}
\begin{tabular}{ccccc}
\hline
\hline
~~~~$x_1$~~~~ & ~~~~$x_2$~~~~  & ~~~~$\int_{x_1}^{x_2} N_h\, dx$~~~~  & ~~~~$\int_{x_1}^{x_2}
x\,N_h\,dx$~~~~ & ~~~~$x_{\rm equivalent}$~~~~ \\ \hline
0.0750 & 1.0000 & 3  & 0.546 & 0.182  \\
0.0350 & 0.0750 & 3  & 0.155 & 0.052 \\
0.0100 & 0.0350 & 9  & 0.167 & 0.018 \\
0.0047 & 0.0100 & 9  & 0.062 & 0.007 \\
0.0010 & 0.0047 & 30 & 0.069 & 0.002 \\
\hline
\hline
\end{tabular}
\end{center}
\end{table}

\begin{figure}
\postscript{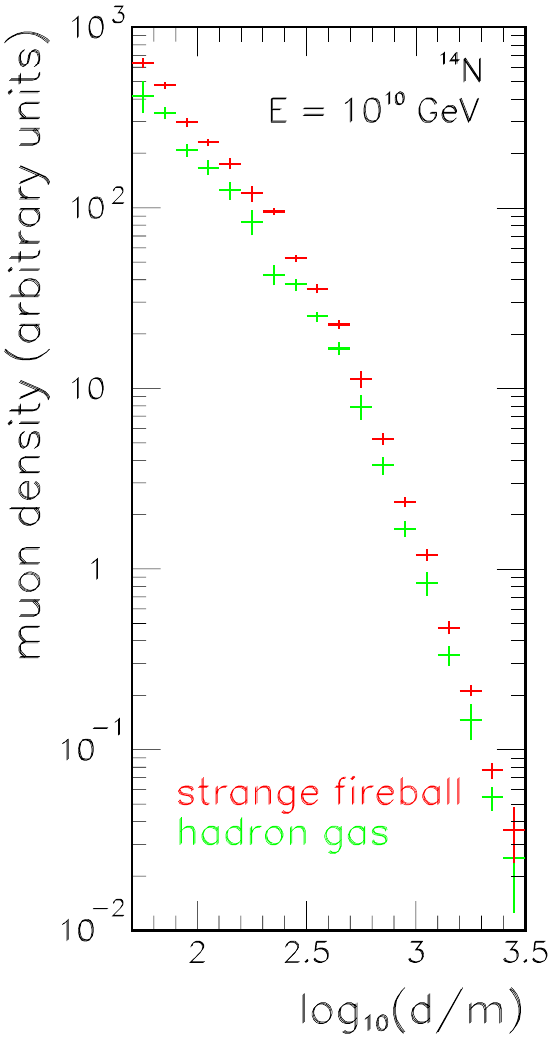}{0.99}
\caption{Density distributions at ground level of $\mu^\pm$ as a
  function of the distance $d$ to the shower axis. The error bars
  indicate the RMS fluctuations.}
\label{fig:2}
\end{figure}

All secondary particles are boosted to the laboratory frame. The
particles are tightly beamed due to their very high boost. The boosted
secondaries are then injected into {\sc aires} as primaries of an air
shower initiated at the collision vertex.  The vertex of the primary
interaction is determined using the mean free path of a $^{14}$N
nucleus with $E = 10^{10}~{\rm GeV}$.  We set the observation point at
1.5~km above sea level, which is the altitude of Auger. All shower
particles with energies above the following thresholds were tracked:
750~keV for gammas, 900~keV for electrons and positrons, 10~MeV for
muons, 60~MeV for mesons, and 120~MeV for nucleons. The geomagnetic
field was set to reproduce that prevailing upon the Auger
experiment. For further details, see Appendix~\ref{App_B}. Secondary
particles of different types in individual showers were sorted
according to their distance $d$ to the shower axis. Our results are
encapsulated in Fig.~\ref{fig:2}, where we compare the density
distribution of $\mu^\pm$ at ground level for a typical nitrogen
shower processed by the {\sc aires} kernel with that of a nitrogen
fireball explosion.  The vertical axis is given in arbitrary units to
indicate the large systematic uncertainty in the normalization of the
lateral distribution, which is induced by the different predictions of
high-energy hadronic event generators.  Importantly, the comparison of
hadron and fireball fluxes is {\it not} arbitrary.  It is easily seen
that the number of muons at $d= 1~{\rm km}$ is about 40\% higher in
the fireball-induced shower.

The third order approximation should include a precise determination
of the fireball particle spectra using scaling hydrodynamic equations,
which contain the probability amplitudes for $s \bar s$ production and
annihilation both in the fireball phase and in the hadron gas
phase~\cite{Rafelski:1982pu,Rafelski:1982ii,Kapusta:1986cb,Matsui:1985eu,Matsui:1986dx,Koch:1986ud}. It
should also contain a thorough high-level competitive analysis of
theoretical systematics emanating from hadronic interaction
models. This rather ambitious project is beyond the scope of this
paper, but will be the topic of a future publication.

\section{Conclusions}
\label{s4}

Teasing out the physics of ultrahigh-energy cosmic rays has proven to
be extraordinarily challenging. The Pierre Auger Observatory employs
several detection methods to extract complementary information about
the extensive air showers produced by ultrahigh-energy cosmic
rays~\cite{ThePierreAuger:2015rma}.  Two types of
instruments are employed: Cherenkov particle detectors on the ground
sample air shower fronts as they arrive at the Earth's surface,
whereas fluorescence telescopes measure the light produced by air-shower 
particles exciting atmospheric nitrogen.  These two detector
systems provide complementary information, as the surface detector
(SD) measures the lateral distribution and time structure of shower
particles arriving at the ground, and the fluorescence detector (FD)
measures the longitudinal development of the shower in the
atmosphere. A subset of {\it hybrid} showers is observed
simultaneously by the SD and FD. These are very precisely measured and
provide an invaluable tool for energy calibration, minimizing
systematic uncertainties and studying composition by simultaneously
using SD and FD information. Very recently, the Pierre Auger
Collaboration exploited the information in individual hybrid events
initiated by cosmic rays with $10^{9.8} \alt E/{\rm GeV}
\alt 10^{10.2}$ to study hadronic interactions at ultrahigh
energies~\cite{Aab:2016hkv}.  The analysis indicates that the observed
hadronic signal of these showers is significantly larger (30 to 60\%) than
predicted by the leading LHC-tuned models, with a corresponding excess
of muons.  The significance of the discrepancy between data (411
hybrid events) and model prediction is above about  $2.1
\sigma$. A deployment of a $4~{\rm m}^2$ scintillator on top of each
SD is foreseen as a part of the AugerPrime upgrade of the Observatory
to  measure the muon and electromagnetic contributions to the ground
signal separately~\cite{Aab:2016vlz}.  This will provide additional
information to reduce systematic uncertainties and perhaps increase
the significance of the muon excess.

Even though the excess is not statistically significant yet, it is
interesting to entertain the possibility that it corresponds to a real
signal of QCD dynamics flagging the onset of deconfinement.  In this
paper, we have proposed a model that can explain the observed excess in
the muon signal.  We have assumed that ultrarelativistic nuclei ($E
\agt 10^{9.8}~{\rm GeV}$) that collide in the upper atmosphere could
create a deconfined thermal fireball which undergoes a sudden
hadronization.  At production, the fireball has a very high matter
density and consists of gluons and two flavors of light quarks
($u,d$). Because the fireball is formed in the baryon-rich projectile
fragmentation region, the high baryochemical potential damps the
production of $u \bar u$ and $d \bar d$ pairs, resulting in gluon
fragmentation mainly into $s \bar s$. The strange quarks then become
much more abundant and upon hadronization the relative density of
strange hadrons is significantly enhanced over that resulting from a
hadron gas. We have shown that the augmented production of strange
hadrons by the fireball, over that resulting from a hadron gas alone,
provides a mechanism to increase the muon content in atmospheric
cascades by about 40\%, in agreement with the data of the Auger
facility.

Contrary to previous proposals~\cite{Farrar:2013sfa,Allen:2013hfa,AlvarezMuniz:2012dd} 
to explain the muon excess in Auger data our model relies on the assumption
that ultrahigh-energy cosmic rays are heavy (or medium mass)
nuclei. As noted elsewhere~\cite{Ahlers:2009rf}, upper limits on the
cosmic diffuse neutrino flux provide a constraint on the proton
fraction in ultrahigh-energy cosmic rays, and therefore can be used to
set indirect constraints on the model proposed herein. In particular,
the nearly guaranteed flux of cosmogenic neutrinos is a decay product
from the generated pions in interactions of ultrahigh energy cosmic
rays with the cosmic microwave
background and related radiation~\cite{Beresinsky:1969qj}. 
The spectral shape and intensity
of this flux depend on whether the cosmic-ray particles are protons or
heavy nuclei.  For proton primaries, the energy-squared-weighted flux
peaks between $10^{9.6}$ and $10^{10}~{\rm GeV}$, and the intensity is
around 1 in Waxman-Bahcall (WB)
units~\cite{Stecker:1978ah,Hill:1983xs,Engel:2001hd,Fodor:2003ph,Anchordoqui:2007fi,Ahlers:2010fw,Kotera:2010yn,Kampert:2012mx}.\footnote{1~WB
  = ${\rm GeV} \, ({\rm cm}^{2} \, {\rm s} \, {\rm
    sr})^{-1}$~\cite{Waxman:1998yy}.} For heavy nuclei, the peak is at
much lower energy (around $10^{8.7}~{\rm
  GeV}$~\cite{Hooper:2004jc,Ave:2004uj}) and the intensity is about
0.1 to 0.01~WB, depending on source
evolution~\cite{Kotera:2010yn,Kampert:2012mx}. The sensitivity of
existing neutrino-detection facilities has a reach to 1~WB, challenging
cosmic-ray models for which the highest energies are
proton-dominated~\cite{Ahlers:2012rz,Aab:2015kma,Unger:2015laa,Aloisio:2015ega,Heinze:2015hhp,Supanitsky:2016gke,Aartsen:2016ngq,Yoshida:2016hba}. Next-generation neutrino detectors will systematically probe the
entire range of the parameter space of cosmogenic
neutrinos~\cite{Allison:2011wk,Aartsen:2014njl,Martineau-Huynh:2015hae,Adrian-Martinez:2016fdl,Neronov:2016zou,Connolly:2016pqr}. Observation
of the cosmogenic neutrino flux with intensity of 0.1 to 0.01~WB could
become the smoking gun for the ideas discussed in this
paper. Complementary information can be obtained with accompanying
cosmogenic photons~\cite{Hooper:2010ze}. As a matter of fact, the
Pierre Auger Observatory has begun to probe the region of the
parameter space relevant for proton primaries~\cite{Aab:2016agp}. 
A third probe of proton-dominated UHECR models is the extragalactic
gamma-ray background seen by Fermi-LAT (and, in the future, CTA)~\cite{Supanitsky:2016gke,Berezinsky:2016jys}.  For
heavy nuclei, however, the cosmogenic photon intensity is almost
negligible and therefore cannot be used (for the time being) as a
harbinger signal~\cite{Anchordoqui:2006pd}.

For $10^{9.5} \alt E/{\rm GeV} \alt 10^{10.6}$, the mean and
dispersion of $X_{\rm max}$ inferred from fluorescence Auger data
point to a light composition (protons and helium) towards the low end
of this energy bin and to a large light-nuclei content (around helium)
towards the high end (see Fig.~3 in~\cite{Aab:2016zth}). However, when
the signal in the water Cherenkov stations (with sensitivity to both
the electromagnetic and muonic components) is correlated with the
fluorescence data, a light composition made up of only proton and
helium becomes inconsistent with observations~\cite{Aab:2016htd}.  The
hybrid data indicate that intermediate nuclei, with $A \simeq 14$,
must contribute to the energy spectrum in this energy bin.  Moreover,
a potential iron contribution cannot be discarded. 
  
At this stage, it is worthwhile to point out that the production of a
fireball may modify the shower evolution. For example, after emitting
$K^+$ and $K^0$, the fireball has a finite excess of $s$ quarks, and
because of the $s$ quarks' stabilizing effects, the fireball could
form heavy multiquark droplets $S$, with large
strangeness~\cite{Witten:1984rs,Liu:1984ta,Farhi:1984qu,Halzen:1984qc,Greiner:1987tg}.
The energy lost by the $S$ particles during collision with nucleons is
primarily through hard scattering, and so the fractional energy loss
per collision is $\sim {\rm
  GeV}/M_S$~\cite{Anchordoqui:2004bd,Anchordoqui:2007pn}.  $S$
production may thus slow down the shower evolution. Although this
effect has not been included in our simulations, one may wonder
whether the structure observed in the elongation rate above about
$10^{10.4}~{\rm GeV}$~\cite{Aab:2014kda} could be ascribed to the
onset of $S$ production.  Of course one would not expect a fireball to
be created when nuclei just slide along each other. The admixture of
peripheral and fireball collisions would then produce large
fluctuations in the number of muons at ground level. However, since
the critical energy for charged pions $\xi_c^{\pi^\pm}$ and kaons
$\xi_c^{K^\pm}$ is roughly the same, the elongation rate of the muon
channel would be almost unaltered and so the muon shower maximum
$X_{\mu, {\rm max}}$ would have small fluctuations. Moreover, if the
fireball indeed modifies the elongation rate of the electromagnetic
shower, the peripheral collisions would also tend to increase the
dispersion of $X_{\rm max}$, mimicking what is expected for a light
composition in the canonical framework where no fireballs are being
produced in this energy range.

In closing, we comment on the differences between our model and the
proposal by Farrar and Allen (FA)~\cite{Farrar:2013sfa}, which also
relies on QCD dynamics at high temperature.  To better understand the
differences between these models we first note that at low energies,
QCD exhibits two interesting phenomena: confinement and (approximate)
spontaneous chiral-symmetry breaking.  These two phenomena are
strong-coupling effects, invisible to perturbation theory.  The
confinement force couples quarks to form hadrons and the chiral force
binds the collective excitations to Goldstone bosons.  

As a matter of course, there is no relation {\it a priori} between
these two phenomena; in thermodynamics, the associated scales are
characterized by two distinct temperatures, $T_c$ and $T_\chi$.  For
$T> T_c$, hadrons dissolve into quarks and gluons, whereas for $T>
T_\chi$, the chiral symmetry is fully restored and quarks become
massless, forming an ideal quark-gluon plasma.  The $\mu_B - T$ phase
diagram of hadronic matter thus contains a {\it confined phase}
consisting of an interacting gas of hadrons (a {\it resonance gas})
and a {\it deconfined phase} comprising a (non-ideal) gas of quarks
and gluons~\cite{Rapp:1999ej}. The {\it phase boundary} reflects the
present uncertainties from lattice QCD extrapolated to finite
baryochemical potential. The intermediate region in-between the hadron
gas and the ideal quark-gluon plasma is the domain of the thermal
fireball.  The existence of this intermediate region with deconfined
massive quarks and gluons is also conjectured from high-statistics
lattice-QCD calculations, which indicate that, for $T \sim 3 T_c$, the
energy density is about 85\% of the Stefan-Boltzmann energy density
for the ideal quark-gluon plasma~\cite{Bazavov:2009zn}. Note that this
temperature is not inconsistent with our estimate in (\ref{T}).  In
the FA model, the pion suppression is a direct consequence of massless
quarks living above the chiral symmetry restoration temperature, i.e.,
$T > T_\chi$. In our model, however, the pion suppression is the
result of the large baryochemical potential which forbids the creation
of light $u \bar u$ and $d \bar d$ pairs, allowing abundant production
of massive $s \bar s$ via gluon fragmentation. This process naturally
occurs in the fireball boundary phase (where $T < T_\chi$), and is a
consequence of the high nucleon density of Lorentz-boosted nuclei. In
principle, it is possible that the muon excess observed in Auger data
originates in a combination of these two high-temperature QCD
phenomena.  Note that if ultrahigh-energy cosmic rays are heavy (or
medium mass) nuclei and the observed muon excess is not the result of
a large baryochemical potential suppressing the production of $u\bar
u$ and $d \bar d$ pairs, but rather an effect of chiral symmetry
restoration, then the excess should also be visible at $\sqrt{s}
\approx 80~{\rm TeV}$, which corresponds to the center-of-mass energy
in collisions of projectile cosmic ray protons with $E \approx
10^{9.5}~{\rm GeV}$. This disparity could be use to discriminate among
dominance between these two pion-suppression mechanisms.

On the one hand, a model consistent with all data requiring heavy
nuclei at the high-energy end of the spectrum is generally considered
a bit disappointing~\cite{Aloisio:2009sj}, especially for neutrino and
cosmic-ray astronomy. On the other hand, we have shown that even if
heavy nuclei dominate at the highest energies, upon scattering in the
Earth's atmosphere these nuclei could become compelling probes of QCD
dynamics at high temperatures, particularly in the
not-so-well-understood fireball boundary phase. Should this be the
case, future AugerPrime data would provide relevant information on the
strange quark density of the nucleon, complementing measurements at
heavy-ion colliders.

\acknowledgments{We would like to acknowledge many useful discussions
  with our colleagues of the Pierre Auger Collaboration. L.A.A.\ is
 supported by U.S. National Science Foundation (NSF) Grant
 No. PHY-1620661 and by the National Aeronautics and Space
 Administration (NASA) Grant No. NNX13AH52G.  T.J.W.\ is supported in
 part by the Department of Energy (DoE) Grant No.\ DE-SC0011981. }

\appendix

\section{Fermi--Dirac and Bose--Einstein integrals}
\label{App_A}

For the sake of completeness, in this appendix we provide all of the
formulae used for computing the number densities of $\bar{s}$, $\bar
q$, and $g$.

The Fermi-Dirac and Bose-Einstein distributions (\ref{FD-BE}) can be
rewritten as infinite sums of Boltzmann distributions,
\begin{eqnarray}
f(p)_{i,\pm}&=&\frac{1}{(2\pi)^3}\frac{1} {e^{(\sqrt{p^2 + m_i^2} -\mu_i)/T} \pm 1} \nonumber\\
&=&\frac{1}{(2\pi)^3} \sum_{n=1}^\infty (\mp)^{n+1} e^{-n \, (\sqrt{p^2 + m_i^2}-\mu_i)/T} \, .
\end{eqnarray}
Following~\cite{Chaudhuri:2012yt} we introduce the dimensionless variables, $z$ and $\tau$:
\begin{eqnarray}
z =\frac{m_i}{T};  \quad \tau=\frac{\sqrt{p^2+m_i^2}}{T}; \quad
|p|  =T\sqrt{\tau^2-z^2};  \nonumber \\ 
|p| \ d|p|=T^2 \tau d\tau; \quad   
|p|^2 \ d|p|  = T^3 \tau \sqrt{\tau^2-z^2} d\tau .  \nonumber
\end{eqnarray}
In terms of $\tau$ and $z$, the number density for the Boltzmann
distribution can be written as,
 \begin{eqnarray}
n_i^{\rm B} & = & \int \frac{d^3p}{(2 \pi)^3} \, e^{(\mu_i - \sqrt{p^2 + m_i^2})/T
} \nonumber \\
& = &4\pi\frac{T^3} {(2\pi)^3} e^{\mu_i/T} \int_z^\infty
d\tau \ (\tau^2-z^2)^{1/2} \ \tau \ e^{-\tau} \, .
\label{nB}
\end{eqnarray}
 A closed form expression can be found for $n_i^{\rm B}$ in terms of the modified Bessel function 
 \begin{equation} 
K_n(z)=\frac{2^n n!}{(2n)!} \frac{1}{z^n} \int_z^\infty d\tau
(\tau^2-z^2)^{n-1/2}  \ e^{-\tau} \, .
\label{Kn1}
\end{equation} 
 Note that $K_n(z)$ has another representation, which follows from (\ref{Kn1}) by partial integration, \begin{equation}
K_n(z)=\frac{2^{n-1} (n-1)!}{(2n-2)!} \frac{1}{z^n} \int_z^\infty\tau
(\tau^2-z^2)^{n-3/2}  \ \tau \ e^{-\tau}.
\label{Kn2}
\end{equation}  
The modified Bessel function has a recurrence relation, 
\begin{equation}
K_{n+1}(z)=\frac{2nK_n(z)}{z}+K_{n-1}(z) \, ,
\end{equation} 
such that if the expressions for $K_0(z) $ and $K_1 (z)$ are known, all the others can be easily obtained. 
From (\ref{Kn2}) it is straightforward to obtain
 \begin{equation}
K_2(z)= \frac{1}{z^2} \int_z^\infty\tau (\tau^2-z^2)^{1/2} \  \tau \ e^{-\tau},
\end{equation}   
and so (\ref{nB}) becomes
 \begin{eqnarray} 
n_i^{\rm B} &=&\frac{T^3} {2\pi^2}  z^2 K_2(z)=\frac{T^3}{2\pi^2} \left
  (\frac{m_i}{T}\right )^2 \ K_2\left (\frac{m_i}{T}\right ) \ e^{\mu_i/T}
\, .
\end{eqnarray} 
For the Fermi-Dirac distribution, the number density is found to be
\begin{equation} 
n_i^{\rm FD} = \frac{T^3}{2\pi^2} \sum_{n=1}^\infty (-1)^{n+1} \
\frac{1}{n^3} \left (\frac{nm_i}{T}\right )^2 K_2\left
  (\frac{nm_i}{T}\right ) \  e^{n\mu_i/T}  \, .
\label{FD}
\end{equation}
For the Bose-Einstein distribution, the number density is given by
\begin{eqnarray} 
n_i^{\rm BE}  = \frac{T^3}{2\pi^2} \sum_{n=1}^\infty \frac{1}{n^3}
\left (\frac{nm_i}{T}\right )^2 K_2\left (\frac{nm_i}{T}\right ) \
e^{n\mu_i/T} \, .
\end{eqnarray} 
In the limit $m_i\rightarrow 0$ and $\mu_i \rightarrow 0$ we immediately obtain
\begin{equation}
n_i^{\rm BE} = \frac{T^3}{\pi^2}\sum_{n=1}^\infty
\frac{1}{n^3}=\frac{T^3}{\pi^2} \zeta(3) \, ,
\label{BE}
\end{equation}
where $\zeta(x)$ is the Riemann function. 
Finally, using (\ref{FD}) and (\ref{BE}), it is straightforward to obtain (\ref{ns}), (\ref{nqbar}) and (\ref{ng}).

\section{$\bm{\mu_B \leftrightharpoons T}$ connection}
\label{App_ref}

\begin{figure}
\postscript{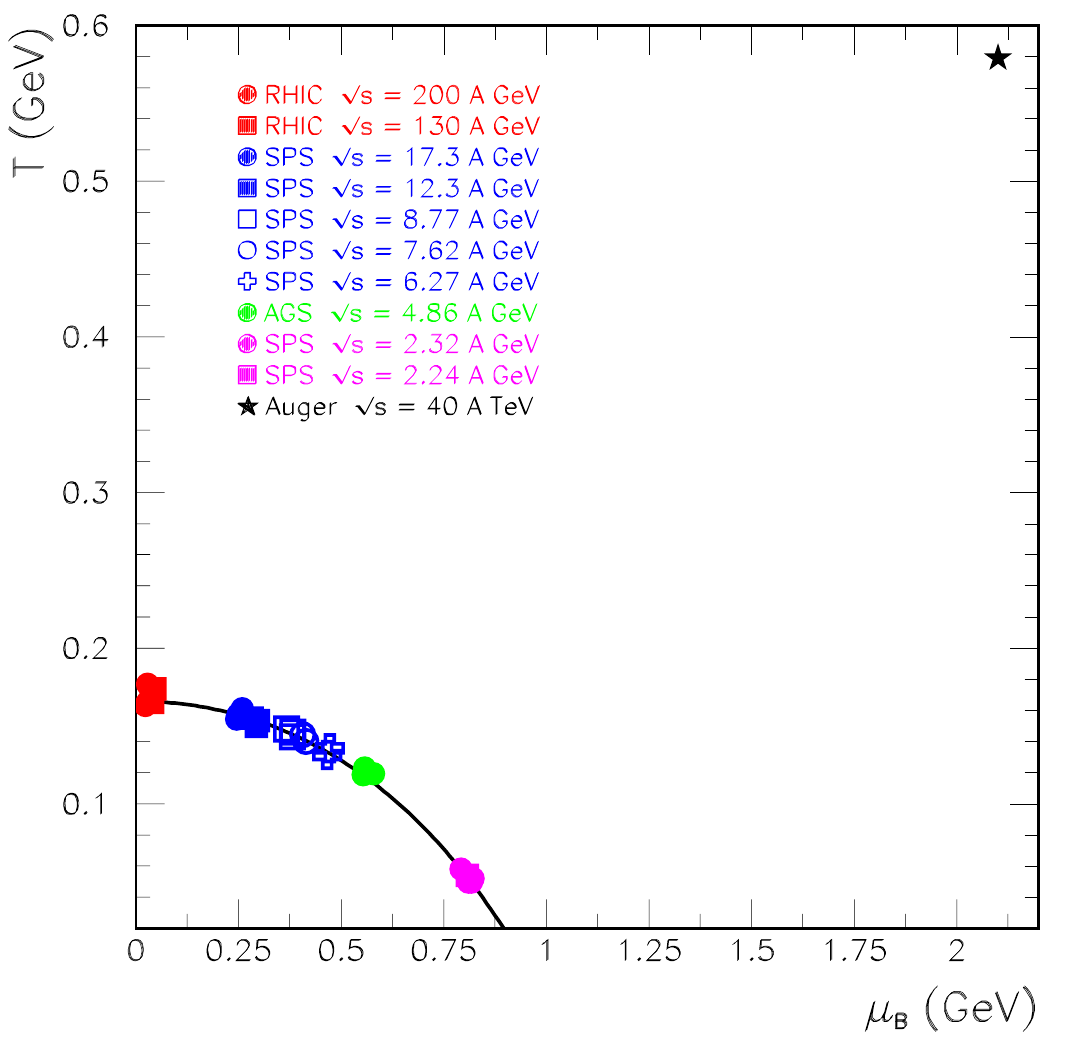}{0.99}
\caption{Relation between $\mu_B$ and $T$ as obtained in statistical-thermal model fits to Au+Au and Pb+Pb collision systems 
by numerous groups~\cite{Cleymans:2004pp,BraunMunzinger:2001ip,Becattini:2003wp,
Becattini:2005xt,Cleymans:1998yb,Redlich:1979bf,Becattini:2000jw,Averbeck:2000sn,Bravina:2002wz,
Adams:2005dq,Florkowski:2001fp,Baran:2003nm}
over a wide range of energies. The solid line is the parameterization (\ref{parametrization2}). 
The lower left hand corner is
dominantly the hadron phase. The region above the crossover line is
dominantly the quark-gluon fireball phase. The black star indicates the required value
for our model to describe Auger data.}
\label{fig:1}
\end{figure}

Our best understanding of the thermodynamical properties of QCD at vanishing baryon density is
rooted on high statistics lattice QCD numerical simulations. However,
if the baryon density is non-zero, these simulations break down. The
only physically relevant analyses for which the obstacles of lattice
QCD can be circumvented are those dealing with small baryon density
or, more accurately, small $\mu_B$. Namely, if we are interested in the
observable $O (\mu_B)$, we can expand it in powers of $\mu_B$ as
\begin{equation}
O(\mu_B) = O_0 + O_1 \, \mu_B + O_2 \, \mu_B^2 + \cdots
\end{equation}
and try to determine the series coefficients~\cite{Bonati:2016fbi}. 
Which values of $\mu_B$ can be considered {\it small enough}  
to give reliable results with this procedure is
something that can only be determined {\it a posteriori}
by the convergence property of the series, limited by  the
accuracy in the evaluation of the expansion coefficients. 
One such observable is the quark-hadron crossover line.
Data from heavy-ion colliders suggest that
the energy dependence of this line can be parametrized as
\begin{equation}
T(\mu_B) = a - b \mu_B^2 - c \mu_B^4 \,, 
\label{parametrization2}
\end{equation}
with $a = (0.166 \pm 0.002)~{\rm GeV}$, $b = (0.139 \pm 0.016)~{\rm
  GeV}^{-1}$ and $c = (0.053 \pm 0.021)~{\rm GeV}^{-3}$, and that
the baryochemical potential can be parametrized as
\begin{equation}
\mu_B(\sqrt{s}) = \frac{a}{1 + b \ \sqrt{s}} \, .
\label{parametrization1}
\end{equation}
with $a= (1.308 \pm 0.028)~{\rm GeV}$ and $b = (0.273\pm 0.008)~{\rm
  GeV}^{-1} $~\cite{Cleymans:2005xv}. Interestingly, the lattice-QCD
calculation converges towards the quark-hadron crossover line as
\mbox{$\mu_B \to 0$~\cite{Becattini:2012xb,Becattini:2016xct},} but it
appears to depart from this line at large values of
$\mu_B$~\cite{Tawfik:2004sw}. This a widely discussed
feature~\cite{Letessier:2005qe,McLerran:2008nn,Andronic:2009gj,Li:2016wzh,McLerran:2016ozl}
which has, however, not been conclusively understood. (Several
hadronization schemes have been proposed, see
e.g.~\cite{Torrieri:2002jp}. They differ in the geometry and in the
flow velocity profile.) This perplexing region may well be relevant to
cosmic-ray observations.  Our results in the region shown in Fig.~\ref{fig:1},
suggest that this is so.

\section{Monte Carlo simulation of air showers}
\label{App_B}

The {\sc aires}
simulation engine~\cite{Sciutto:1999jh}  provides full space-time particle propagation in a
realistic environment, taking into account the characteristics of the
atmospheric density profile (using the standard US atmosphere), the
Earth's curvature, and the geomagnetic field (calculated for the
location of Auger with an uncertainty of a few
percent~\cite{Cillis:1999gk}).  The following particles are tracked in
the Monte Carlo simulation: photons, electrons, positrons, muons,
pions, kaons, eta mesons, lambda baryons, nucleons, antinucleons, and
nuclei up to $Z=36$.  The high-energy collisions are processed invoking
external hadronic event generators, whereas the low-energy ones are
processed using an extension of the Hillas splitting algorithm~\cite{Knapp:2002vs}.  

The {\sc aires} program consists of various interacting procedures
that operate on a data set with a variable number of records.  Several
data arrays (or stacks) are defined. Every record within any of these
stacks is a particle entry and represents a physical particle. The
data contained in every record are related to the characteristics of
the corresponding particle. The particles can move inside a volume
within the atmosphere where the shower takes place. This volume is
limited by the ground, the injection surfaces, and by vertical planes
which limit the region of interest. Before starting the simulation,
all the stacks are empty. The first action is to add the first stack
entry, which corresponds to the primary particle. Then the stack
processing loop begins. The primary is initially located at the
injection surface, and its downwards direction of motion defines the
shower axis. After the primary's fate has been decided, the
corresponding interaction begins to be processed. The latter generally
involves the creation of new particles which are stored in the empty
stacks and remain waiting to be processed. Particles entries are
removed when one of the following events happen: {\it (i)}~the energy
of the particle is below the selected cut energy; {\it (ii)}~the
particle reaches ground level; {\it (iii)}~a particle going upwards
reaches the injection surface; {\it (iv )}~a particle with
quasi-horizontal motion exists the region of interest. After having
scanned all the stacks, it is checked whether or not there are new
particle entries pending further processing. If the answer is
positive, then all the stacks are scanned once more; otherwise the
simulation of the shower is complete.

For the present analysis, we use the {\sc aires}
module for {\it special and/or multiple primary particles}. This
useful feature allows to dynamically call a user-defined module that
tracks the interactions of a bundle of particles returning a handy list
of secondaries, which can be conveniently controlled 
by the propagating engine for further processing.

\end{document}